%% file: procMoriond.tex
\documentclass[11pt]{article}
\usepackage{moriond}
\usepackage{amsmath}
\usepackage{amsfonts}
\usepackage{latexsym}
\usepackage{epsfig}
\newcommand{\be}{\begin{equation}}
\newcommand{\ee}{\end{equation}}
\newcommand{\bea}{\begin{eqnarray}}
\newcommand{\eea}{\end{eqnarray}}

\renewcommand{\d}{\mathrm{d}}

\DeclareMathSymbol{\mg}{\mathrel}{symbols}{"1D}

\newcommand{\ga}{\alpha}

\newcommand{\gd}{\delta}

\newcommand{\gf}{\phi}

\newcommand{\gx}{\xi}
\newcommand{\gm}{\mu}

\newcommand{\gl}{\lambda}

\newcommand{\gp}{\pi}
\newcommand{\gps}{\psi}

\newcommand{\gG}{\Gamma}

\newcommand{\gF}{\Phi}

\newcommand{\gL}{\Lambda}
\newcommand{\gS}{\Sigma}

\newcommand{\cP}{{\cal P}}

\newcommand{\cR}{{\cal R}}

\newcommand{\der}{\partial}

\newcommand{\bit}{\bibitem}

\newcommand{\labl}[1]{\label{#1}}
\newcommand{\half}{\frac 12 }
\newcommand{\shalf}{{\scriptstyle \half}}

\newcommand{\beq}{\begin{equation}}
\newcommand{\eeq}{\end{equation}}
\newcommand{\barr}{\begin{array}}
\newcommand{\earr}{\end{array}}
\newcommand{\equ}[1]{\begin{gather} #1 \end{gather}}

\newcounter{oldcounter}

\newcommand{\bgf}{{\bar\phi}}

\newcommand{\Intr}{\mathbb{Z}}

\begin{document}

\vspace*{4cm}
\title{Higgs mass sensitivity and localization due to Fayet-Iliopoulos terms
on 5D orbifolds\footnote{Invited talk given at the XXXVIIth
 Rencontres de Moriond session devoted to Electroweak Interactions And Unified
 Theories, March 9-16 2002, Les Arcs (France).}}

\author{Stefan Groot Nibbelink }

\address{
Physikalisches Institut der Universitat Bonn, \\
Nussallee 12, 53115 Bonn, Germany,\\
nibblink@th.physik.uni-bonn.de.
}

\maketitle\abstracts{
In this talk we review calculations of FI--tadpoles in 5 dimensional 
(non--)supersymmetric orbifold theories. Some consequences of these
tadpoles are discussed: quadratic Higgs--mass sensitivity to a high
scale, and localization of bulk matter fields to the orbifold fixed points. 
}

\section{Introduction}

Models with 5 dimensional global supersymmetry compactified on 
orbifolds may be good candidates for extensions of the standard model and 
have interesting phenomenological applications. The orbifolds we consider
in this talk can have both a supersymmetric ($S^1/\Intr_2$) as well
as a non--supersymmetric ($S^1/\Intr_2 \times \Intr_2'$) spectrum.

The underlying 5 dimensional $N=1$ supersymmetry can give rise to many
impressive ultra--violet properties while the orbifold compactification
can produce phenomenologically interesting particle spectra. 
Let us mention a particular intriguing model proposed by
Barbieri, Hall, Nomura (BHN) \cite{Barbieri}, which has some remarkable
features:  Although this model has the low energy spectrum identical to the
standard model it is constructed from a supersymmetric theory with
vector and hyper multiplets compactified on the orbifold 
$S^1/\Intr_2 \!\times\! \Intr_2'$. 
In the following table the Kaluza--Klein spectrum of this model is
presented. 
\begin{center}
\vspace{5mm}
\scalebox{.75}{\mbox{
\includegraphics*[40mm,210mm][165mm,280mm]{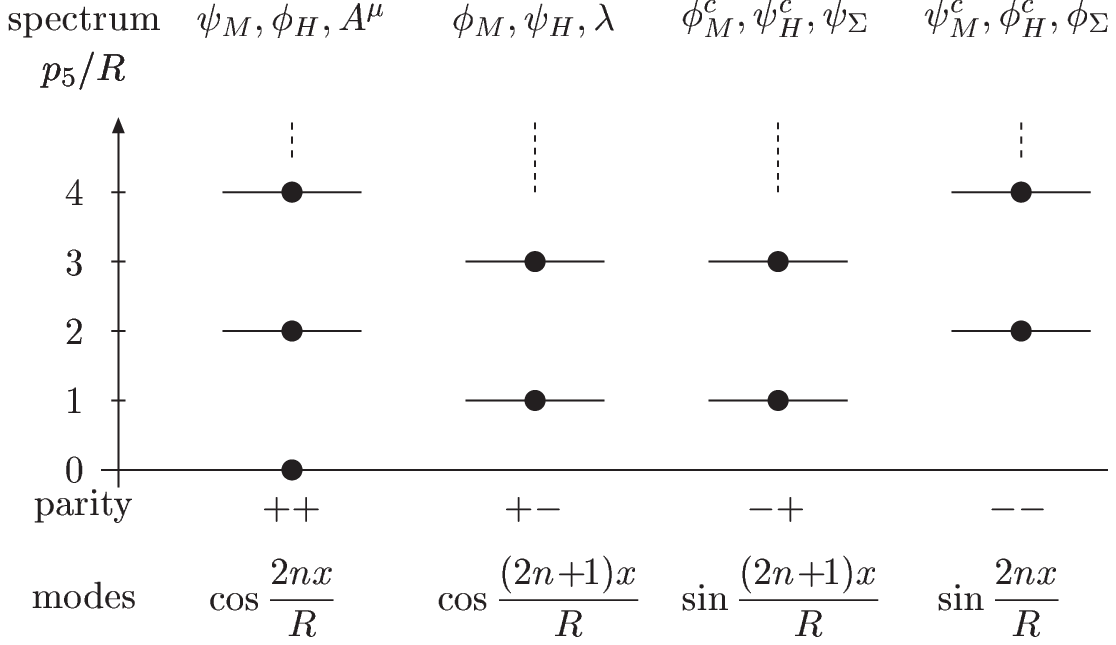}
}}
\vspace{5mm}
\end{center}
The parity assignment dictates the mode expansion of a given field.
The field content of this model consists of a complex Higgs
scalar $\gf_H$, its Higgsino $\gps_H$, the standard model fermions
$\gps_M$,  their mirrors $\gps_M^c$, and the sfermions $\gf_M,
\gf_M^c$ that form 5 dimensional hyper multiplets. 
Whereas, the standard model gauge fields $A_\gm, A_5$, the two gauginos $\gl,
\gps_\gS$ and real scalar $\gF$ form a vector multiplet. 
The 5th component of the gauge field $A_5$ in 5 dimension
and the real scalar $\gF$ reside in $\gf_\gS$. 
All these are all functions of the 5th dimension of with radius $R$. 

In this proceedings we consider two stability issues of such models
in 5 dimensions that are consequences of divergent
FI--terms: 1) Higgs mass sensitivity to the cut--off, and 2)
localization of charged bulk matter. Before discussing these issues in the
following sections, let us introduce them briefly here. 

In the recent literature these types of orbifold models were claimed
to  have an extremely mild ultra--violet (UV) behavior \cite{finiteresults}:
the effective potential was claimed to be finite at one loop or even to
all orders in perturbation theory. Others \cite{Ghilencea}
raised objections to such claims in the case of models that do not
possess any global supersymmetry, that may provide an obvious 
UV--protection for the Higgs mass. It turned out that, like in 4 dimensional
supersymmetric models, Fayet--Iliopoulos tadpole may introduce a
quadratic divergence.  

The second issue boils down to the question whether any configuration
of brane and bulk fields is stable. 
Charged bulk fields can become strongly localized due to the
effect of FI--terms in 5 dimensions induced at one loop. 
If this happens the original setup was not stable under quantum
corrections and should therefore not be considered as the appropriate
starting point for perturbative calculations.

\section{The zero mode Fayet--Iliopoulos term}

In supersymmetric field theory in 
4 dimensions the FI--term is either quadraticly divergent or vanishes 
at one loop.  In the following we focus on the Higgs sector of the
BHN--model to discuss the effect of the zero mode FI--term in the
effective 4 dimensional theory. The diagram of the FI--contribution to
the selfenergy of a scalar is given by:
\begin{center}
\vspace{5mm}
\input{FI.pstex_t}
\vspace{5mm}
\end{center}
The dotted line corresponds to the auxiliary field $D^\parallel$ 
of the Abelian gauge multiplet in 4 dimensions. (The notation
$D^\parallel$ indicates that this the component of the triplet of
auxiliary fields of the vector multiplet that has a KK zero mode after
the orbifolding.) In ref.\ \cite{SHPD} we have
investigated what happens to the FI--term in the effective field theory 
coming from 5 dimensions with a mass spectrum of the complex scalars 
of the hyper multiplet on $S^1/\Intr_2 \times \Intr_2'$. 
We denote the charges of the even and odd KK scalars 
by $q^{++}_n = - q^{--}_n = 1$. Formally, the expression for 
the one loop contribution to the FI term reads 
\equ{
\gx_0^{} = \sum_{n, \ga} \,g q^{\ga\ga}_n \int \frac{\d^4 p_4}{(2\gp)^4} 
\frac {1}{p_4^2 + ({m_n^{\ga\ga}})^2 + m^2},
\labl{FIoneloopKK}
} 
where $m_n^{\ga\ga} = 2n/R$ and the sum for $\ga = +$ is over 
$n \geq 0$, while for $\ga = -$ over $n > 0$. 
In order to be able to calculate this quantity in a rigorous way we 
employ dimensional regularization of a compact dimension 
introduced in ref.\ \cite{sgn} 
\equ{
\gx_0^{} = g\,  \int \frac{\d^{D_4} p_4}{(2\gp)^{D_4}} 
\int_\ominus \frac { \d^{D_5} p_5}{2\gp i}
\left\{
\frac{\cP^{++}(p_5)}{p_4^2 + p_5^2 + m^2} - 
\frac{\cP^{--}(p_5)}{p_4^2 + p_5^2 + m^2}
\right\}.
}
These integrals are defined as complex functions of the dimensions
$D_4$ and $D_5$ by  
\equ{
\int_\ominus \d^{D_5} p_5 \int \d^{D_4} p_4\, 
 \equiv  
 \int_{\ominus} \frac{\d p_5}{2\gp i}\int_0^\infty \d p_4 
\, \cR_{4}( p_4) \cR_{5}(p_5) \, 
}
with the regulator functions $\cR_{4}(p_4)$ and $\cR_5(p_5)$ given by 
\equ{
\cR_{4}(p_4) =  
\frac {2 \gp^{\half D_4}}{\gG(\half D_4)}\, p_4^3 \, 
\Bigl( \frac {p_4}{\gm_4} \Bigr)^{D_4 -4}, 
\qquad 
\cR_{5}(p_5) =  
\frac {\gp^{\half D_5}}{\gG(\half D_5)}\, 
\Bigl( \frac {p_5}{\gm_5} \Bigr)^{D_5 -1}.
\labl{regufunctions}
}
With $\ominus$ the contour integration is denoted over the upper and
lower half plane with an anti--clockwise orientation \cite{Arkani,Mirabelli}. 
Substituting the expressions of the pole functions 
\equ{
{\cP^{\pm\pm}} = {\half}  
{\left(
{\pm\frac {1}{p_5}} + \frac {\shalf \gp R}{\tan \shalf \gp R p_5} 
\right)}, 
\labl{polefunS22}
}
gives exactly the same result as the regulated FI term for one
massless complex scalar: 
\equ{
\gx_0^{} = g\,   \int \frac{\d^{D_4} p_4}{(2\gp)^{D_4}} 
\int_\ominus  \frac {\d^{D_5} p_5}{2\gp i} 
\frac 1{p_5} 
\frac {1}{p_4^2 + p_5^2 + m^2} 
= g\, 
\int \frac{\d^{D_4} p_4}{(2\gp)^{D_4}} \frac {1}{p_4^2 + m^2}.  
\labl{FItermzero}
}
Since it behaves as a single particle contribution we can safely take 
$D_5 =1$ giving the 4 dimensional quadratically divergent expression. 

In ref.\ \cite{SHPD} we have shown that the other gauge
contributions give a finite correction and can therefore
never cancel this quadratic divergence. According to ref.\
\cite{Barbieri:2001cz} the correction to the Higgs mass due to this
quadratically divergent FI--tadpole is relatively
small if the cut--off (used to regulate the divergent integral) is taken to be
around $5/R$. The motivation for this value of the cut--off is that
beyond this value the power--running of the gauge couplings
explodes. However, since in principle these are two different types of
``cut--offs'' (one is a regulator while the other corresponds to the
scale of the gauge coupling Landau pole), it is not clear why they
should simply be equal.  
The cut--off at $5/R$ corresponds to a numerical value of a few TeV. 
Of course, with a cut--off of this order the standard model does not
require any fine--tuning in its Higgs sector and neither supersymmetry
nor extra dimensions are required.

\section{Localization of and due to Fayet--Iliopoulos terms}

FI--terms in 5 dimensions do not only affect scalar masses, but they
may also have important consequences for stability of such theories. 
In order not to complicate our discussion here, we 
consider supersymmetric compactification on the orbifold
$S^1/\Intr_2$, with a $U(1)$ vector multiplet $(A_M, \gF, \gl)$ and 
charged hyper multiplets $(\gf_+, \gf_-, \gps)$ in the bulk.

The profile of FI--terms over the 5th dimension is rather intriguing:
as observed in ref.\ \cite{Barbieri:2001cz,Scrucca:2001eb}
the tadpole for $D^\parallel$ leads to a FI--parameter 
\equ{
\gx_{bulk}(x) = \frac{g}{2}
\left(
\frac{\gL^2}{16 \gp^2} +
\frac{\ln \gL^2}{16 \gp^2} \frac 14 \der_x^2
\right)
\Bigl[ \gd(x) + \gd(x - \gp R) \Bigr].
}
The leading quadratic divergence is localized at the two branes as is
signified by the delta--functions. The sub--leading logarithmic
divergence is proportional to the second derivate of these
delta--functions. Similar tadpoles arise for the derivative of the
physical scalar $\gF$ in the gauge multiplet, due to a fermion
(hyperino) loop. \cite{GrootNibbelink:2002wv}  In the
picture below we give the diagrams for both the $D^\parallel$ and the
$\der_x \gF$ tadpole:
\begin{center}
\vspace{5mm}
\scalebox{.9}{\mbox{\input{FIbulk.pstex_t}}}
\vspace{5mm}
\end{center}
The combination $D^\parallel - \der_x \gF$ for these tadpoles is required by
the remaining supersymmetry after compactification on the orbifold 
$S^1/\Intr_2$.\cite{Mirabelli} 

The consequences of this shape of the FI--terms have been investigated
in detail in ref.\ \cite{GrootNibbelink:2002qp}: 
the terms with double derivative on the delta--function, lead either to
delta--like localization to or repulsion from the branes of charge bulk
fields. The reason for this is the non--trivial background profile of
the physical scalar $\gF$ due to its FI--tadpoles, 
affects the shape of the zero mode of the bulk matter fields: 
\equ{
\gf_{0+}(x) = \exp\left\{  g \int_0^x \d x\,   \gF   \right\} 
\bgf_{0+}.
\labl{zeroMode}
}
This effect
may be interpreted as a signal that one has started with a model with
a distribution of the matter fields over the 5th dimension that is
unstable under quantum corrections. Therefore, only models that do not
have this type of  instability should be considered as valid starting
points for detailed phenomenological studies. 

Another important (and related) requirement is, of course, gauge
anomaly cancellation studied in refs.\ 
\cite{Arkani-Hamed:2001is,Pilo:2002hu,Barbieri:2002ic}. In addition in ref.\
\cite{GrootNibbelink:2002qp}  the issue of a parity anomaly on
$S^1$ is raised that can make an orbifold model ill--defined.

\section{Conclusion}

In this talk we have discussed two types of instabilities that can
arise due to Fayet--Iliopoulos terms in 5 dimensional (supersymmetric)
orbifold theories. In the non--supersymmetric BHN--model the
quadratical divergence leads to a quadratic sensitivity of
the Higgs mass to the cut--off. 
The FI--terms have a profile over the 5th dimension 
proportional to delta--functions localized at both boundaries and
second derivatives of those delta--functions. 
This often leads to strong localization of the zero modes of charged bulk
fields, which signals
an instability in the initial distribution of matter over the 5
dimensional bulk and the 4 dimensional boundaries.

\section*{Acknowledgments} 

It is a pleasure to thank D.\ Ghilencea,  H.P.\ Nilles and M.\
Olechowski for the stimulating collaboration during various stages of
the work reported in this proceedings. 
This work is supported by priority grant 1096 of the Deutsche 
Forschungsgemeinschaft and European Commission RTN 
programmes HPRN-CT-2000-00131 / 00148 and 00152.

\end{document}

%% file: FI.pstex_t
\begin{picture}(0,0)%
\includegraphics{FI.pstex}%
\end{picture}%
\setlength{\unitlength}{2763sp}%
\begingroup\makeatletter\ifx\SetFigFont\undefined%
\gdef\SetFigFont#1#2#3#4#5{%
  \reset@font\fontsize{#1}{#2pt}%
  \fontfamily{#3}\fontseries{#4}\fontshape{#5}%
  \selectfont}%
\fi\endgroup%
\begin{picture}(3024,1519)(2089,-5773)
\put(4126,-4561){\makebox(0,0)[lb]{\smash{\SetFigFont{11}{13.2}{\familydefault}{\mddefault}{\updefault}
$h$
}}}
\put(3226,-5536){\makebox(0,0)[lb]{\smash{\SetFigFont{11}{13.2}{\familydefault}{\mddefault}{\updefault}
$D^\parallel$
}}}
\end{picture}

%% file: FIbulk.pstex_t
\begin{picture}(0,0)%
\includegraphics{FIbulk.pstex}%
\end{picture}%
\setlength{\unitlength}{2763sp}%
\begingroup\makeatletter\ifx\SetFigFont\undefined%
\gdef\SetFigFont#1#2#3#4#5{%
  \reset@font\fontsize{#1}{#2pt}%
  \fontfamily{#3}\fontseries{#4}\fontshape{#5}%
  \selectfont}%
\fi\endgroup%
\begin{picture}(2866,1854)(3818,-5773)
\put(5851,-5536){\makebox(0,0)[lb]{\smash{\SetFigFont{12}{14.4}{\familydefault}{\mddefault}{\updefault}
$\Phi$
}}}
\put(3901,-4111){\makebox(0,0)[lb]{\smash{\SetFigFont{12}{14.4}{\familydefault}{\mddefault}{\updefault}
$\phi_+$, $\phi_-$
}}}
\put(3826,-5536){\makebox(0,0)[lb]{\smash{\SetFigFont{12}{14.4}{\familydefault}{\mddefault}{\updefault}
$D^\parallel$
}}}
\put(6151,-4111){\makebox(0,0)[lb]{\smash{\SetFigFont{12}{14.4}{\familydefault}{\mddefault}{\updefault}
$\psi$
}}}
\end{picture}